\documentclass[conference]{IEEEtran}
\IEEEoverridecommandlockouts
\usepackage{cite}
\usepackage{amsmath,amssymb,amsfonts}
\usepackage{algorithmic}
\usepackage{graphicx}
\usepackage{textcomp}
\usepackage{xcolor}
\usepackage{multirow}
\usepackage{hyperref}
\def\BibTeX{{\rm B\kern-.05em{\sc i\kern-.025em b}\kern-.08em
    T\kern-.1667em\lower.7ex\hbox{E}\kern-.125emX}}
\begin{document}

\title{Exploring Latent Space for Generating Peptide Analogs Using Protein Language Models}

\author{
    \IEEEauthorblockN{1\textsuperscript{st} Po-Yu Liang}
    \IEEEauthorblockA{
        \textit{Department of Computer Science} \\
        \textit{University of Cincinnati}\\
        Ohio, United States \\
        liangpu@mail.uc.com
    }\and
    
    \IEEEauthorblockN{2\textsuperscript{nd} Xueting Huang}
    \IEEEauthorblockA{\
        \textit{Department of Pharmaceutical Sciences} \\
        \textit{University of Connecticut, Storrs}\\
        Connecticut, United States \\
        xueting.huang@uconn.edu
    }\and
    
    \IEEEauthorblockN{3\textsuperscript{rd} Tibo Duran}
    \IEEEauthorblockA{
        \textit{Department of Pharmaceutical Sciences} \\
        \textit{University of Connecticut, Storrs}\\
        Connecticut, United States \\
        tibo.duran@uconn.edu
    }\and
    
    \IEEEauthorblockN{4\textsuperscript{th} Andrew J. Wiemer}
    \IEEEauthorblockA{
        \textit{Department of Pharmaceutical Sciences} \\
        \textit{Institute for Systems Genomics} \\
        \textit{University of Connecticut, Storrs}\\
        Connecticut, United States \\
        andrew.wiemer@uconn.edu
    }\and
    
    \IEEEauthorblockN{5\textsuperscript{th} Jun Bai}
    \IEEEauthorblockA{
        \textit{Department of Computer Science} \\
        \textit{University of Cincinnati}\\
        Ohio, United States \\
        baiju@ucmail.uc.edu
    }
}

\maketitle

\begin{abstract}
Generating peptides with desired properties is crucial for drug discovery and biotechnology. Traditional sequence-based and structure-based methods often require extensive datasets, which limits their effectiveness. In this study, we proposed a novel method that utilized autoencoder shaped models to explore the protein embedding space, and generate novel peptide analogs by leveraging protein language models. The proposed method requires only a single sequence of interest, avoiding the need for large datasets. Our results show significant improvements over baseline models in similarity indicators of peptide structures, descriptors and bioactivities. The proposed method validated through Molecular Dynamics simulations on TIGIT inhibitors, demonstrates that our method produces peptide analogs with similar yet distinct properties, highlighting its potential to enhance peptide screening processes. 

\end{abstract}


\section{Introduction}
Biologists often seek peptides for various purposes, such as aiding in the comprehension of biological mechanisms and addressing societal challenges in healthcare. Immunotherapy, for example, relies on immune checkpoint inhibitors to block checkpoint proteins from binding with target proteins, thereby enhancing immune cell activity against cancer cells. The process of peptide discovery, traditionally laborious, has seen acceleration in recent years due to the fast development of computational methods\cite{valentinuzzi2020computational,duran2024might}. These computational methods, including virtual screening, molecular docking, machine learning/deep learning, have revolutionized peptide discovery by various functionalities such as predicting peptide-protein interactions, optimizing peptide sequence, and identifying novel peptide candidates \cite{duran2022molecular,lei2021deep,duran2021molecular}. Among these methods, deep learning stands out as the most advanced computational method \cite{bai2021applying, mamoshina2016applications,lecun2015deep} which focus on reduced part of the vast compound space related to peptides, allowing for the prediction of peptides with a certain degree of accuracy. By leveraging large datasets and sophisticated neural network architectures, the advanced models can uncover bioactive peptides for various applications. Deep learning algorithms, such as various data structured discrimination models \cite{tropsha2024integrating,ghasemi2018neural} and generative models \cite{lin2020relevant,lin2022novo}, have shown remarkable success identifying an optimizing peptides with desired properties. This success has paved the way for the research focused on the generation of peptides sequence with specific properties.

Generating peptides sequence with specific properties recently become a key focus in computer-aided peptide and drug discovery\cite{sharma2023peptide}. Researchers are concentrating on two main approaches: sequence-based method and structure-based method. The sequence-based methods \cite{gupta2019feedback, greener2018design} learn patterns from existing peptides to predict new ones. These methods often heavily rely on known dataset with desired properties or bioactivities. 
One significant challenge of sequence-based methods is determine the desired properties for new peptide. In other words, the limited availability of known peptide sequences with these desired properties can reduce the effectiveness of these methods.
The structure-based models \cite{goverde2023novo, goverde2024computational} generates peptides with known three-dimensional structures. These methods require known structures of peptides with target receptors, which can be difficult to obtain, especially for peptides with desired properties or bioactivities. Additionally, structure-based methods face limitations when dealing with peptides that have disordered or unstable structures.

Our study addresses the challenges of generating peptides analogs without relying on large datasets and structures. We proposed a novel method to generates peptide analogs by exploring the embedding space, requiring only a single peptide sequence. Our proposed method could streamline the peptide discovery process, making it more efficient and less resource-intensive. Leveraging pre-trained protein language models, our method eliminates the need for extensive datasets of similar sequences. We validated the effectiveness of our method using various similarity indicators and performed Molecular Dynamics (MD) simulations on TIGIT inhibitor peptides identified through wet lab experiments. The code used for this research is available at: \url{https://github.com/LabJunBMI/Latent-Space-Peptide-Analogues-Generation}

\section{Related Research}

\subsection{Lab Experiment Based Method}\label{sec:lab_experiment_method}
A variety of approaches can be used to identify and optimize peptide-based inhibitors. Rational design is a classic approach to develop peptide analogs based on identification of key residues contributing to the natural ligand:target interaction. While it allows precise modifications, a limitation is that it requires sequence and/or structure information of both the target protein and its ligand\cite{yin2021rational}. Obtaining this information requires significant time and effort, including extensive mutagenesis scanning and/or generation of a crystal structure. Even with structural data available, only a limited number of peptide analogs can be generated by rational design, potentially missing optimal sequences. Phage display is a powerful tool that enables high-throughput screening of peptides by virtue of a DNA-encoded phage-displayed library and rapid phage amplification\cite{hamzeh2013phage}. However, binding affinity and function of the selected peptides are not guaranteed, requiring further lead optimization to improve the target selectivity, potency and efficacy. Directed evolution is an advanced method for evolving peptide sequences by generating a library using random mutations and conducting iterative cycles of mutations and selections\cite{packer2015methods}. Although it explores a larger sequence space\cite{romero2009exploring}, it requires resources for screening, and its success rate is dependent upon the initial library. Moreover, it can be time-consuming due to the iterative mutations and screenings.

\subsection{Deep Learning Based Method}

In the past decade, numbers of researches studied generating peptides with desired properties using deep learning methods. Some studies focus on the amino acid sequences of proteins. Greener et al. \cite{greener2018design} utilized a conditional variational autoencoder to generate metalloproteins based on a known dataset of amino acid sequences. Gupta et al. \cite{gupta2019feedback} applied generative adversarial networks combined with a pre-trained sequence function prediction model to generate DNA sequences of proteins with desired properties. Biswas et al. \cite{biswas2021low} used evolutionarily related homolog sequences, retrieved using a hidden markov model, to fine-tune an embedding model, which was then employed to predict functional characteristics. They subsequently combined this functional prediction model with the Markov chain Monte Carlo (MCMC) method to generate similar proteins. Goverde et al. \cite{goverde2023novo} attempted to search for sequences with desired three-dimensional structures by inverting the AlphaFold2 structure prediction network \cite{jumper2021highly} through the MCMC method. They further extended this research to validate the effectiveness of their method, with some modifications, on membrane proteins \cite{goverde2024computational}. 

\section{Method}
In this research, we proposed a new method to generate peptide sequences by exploring the protein latent space. 

\textit{Hypothesis} Our study posits that peptides with similar embedding are likely to share higher property similarities, even if their sequence expressions differ. This hypothesis is inspired by the field of word embedding studies, where vector abstract feature representations learned from deep learning model capture semantic meaning \cite{asudani2023word_embedding_impact}. In our research, we proposed that the vector abstract feature representations derived from data distributions encapsulate the bioactivety implications of peptide sequences. By utilizing computational techniques similar to those used in word embedding methodologies, we aim to elucidate the structural and functional relationships between peptides, facilitating the generation of peptide analog sequences without the need for extensive datasets by exploring the protein latent space.

\textit{Definition} Our proposed method employs an autoencoder shaped model to learn the feature embedding. We define our dataset as $X=\{x_0,...,x_i,...,x_n\}$, where $x_i$ is the amino acid sequence of a protein. We define our method as $\hat{y}_{\tau} = g(f(x_i)+\delta_{\tau})$, where $\hat{y}_{\tau} $ is the generated amino acid sequence of protein analog at step $\tau$, function $f(\cdot)$ is a model projecting a protein sequence into the latent space, $\delta_{\tau}$ represents the noise added to the protein embedding  at step $\tau$, and $g(\cdot)$ project the noised embedding back to the sequence.
\begin{figure}[htbp]
    \centering
    \includegraphics[width=\columnwidth]{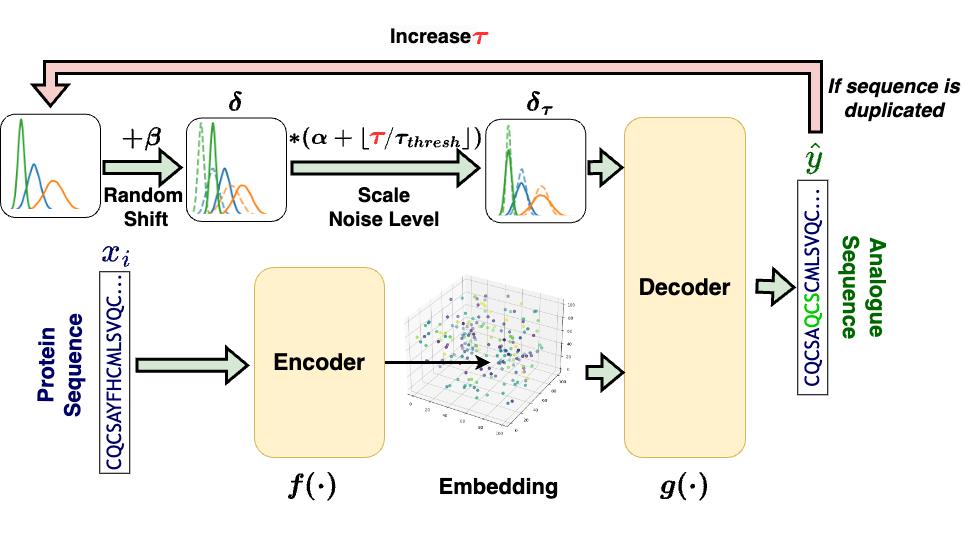}
    \caption{Method Flow Chart}
    \label{fig:method_flow}
\end{figure}
\subsection{Overview of the Proposed Method}

The proposal method overall structure is showing in the Figure \ref{fig:method_flow}. 
First, the embedding step projects peptide sequences from a discrete space into a continuous latent space. This transformation is crucial as it allows us to manipulate the sequences in a more flexible and informative way. Second, we explore the latent space by introducing noise into the embeddings. This step is performed in a systematic manner, starting with lower levels of noise and gradually increasing to higher levels. By doing so, we can generate sequences that are progressively different from the original, enabling us to discover a wide range of similar peptide analogs. Finally, in the decoding step, the noised embeddings are processed by a decoder, which converts the embeddings back into peptide sequences. This step is essential to ensure that the manipulated embeddings are interpretable and usable.

To validate our method, we utilized two embedding models: ProtT5\cite{elnaggar2021prottrans} and ESM\cite{lin2023evolutionary}. Both embedding models, ProtT5 and ESM-2, are transformer-based and incorporate layer normalization modules that standardize hidden states layer-by-layer to enhance model robustness to noise. ProtT5 is an encoder-decoder based embedding model, which allows us to directly use its decoder module for the final step. On the other hand, ESM is a mask-based embedding model, requiring us to train a new decoder model to transform the embeddings back into sequences. This dual-model approach enables us to demonstrate the robustness and versatility of our method across different embedding architectures.

\subsection{Embedding}
In our method, we utilized two state-of-the-art models, as peptide sequence projection function $f(\cdot)$, to embed peptide sequences: ProtT5 and ESM-2. For both models, we use pre-trained versions to leverage their advanced capabilities in understanding protein sequences. We employed token-level embedding instead of sequence-level embedding for both models, where each token is obtained using one-hot encoding of amino acid. This approach allows the decoder module to reverse the embedding back into sequences effectively, ensuring a seamless transition from embedding to decoding.

\subsubsection{ProtT5 Embedding}
We employed the "Prot-T5-XL-Ur50" as ProtT5 model, which is trained on the UniRef50 dataset\cite{suzek2015uniref}. This model provides an embedding size of $1024$ to capture features of peptide sequences. In the ProtT5 model, the first embedding corresponds to the initial \texttt{<\textbackslash{}pad>} token, and the last embedding corresponds to the end token \texttt{<\textbackslash{}s>}. Therefore, we removed these two embedding to focus on the meaningful representations of the peptide sequences.

\subsubsection{ESM-2 Embedding}
We utilized the ESM-2 models's version with $150$ million parameters to match the ProtT5\cite{esm}, making it a fair comparison. In the ESM-2 model, we used the output from the last layer of the sequence module as the peptide embeddings, which have an output size of $640$. This layer captures the  features of the peptide sequences, providing a foundation for the subsequent steps in our method. By incorporating these two different types of embedding models --- encoder-decoder based (ProtT5) and mask-based (ESM-2) --- we aim to evaluate the performance and versatility of our method across various embedding architectures. This dual-model approach ensures a comprehensive understanding of how our method functions with different embedding strategies.

\subsection{Noise}
In this step, we introduce noise $\delta$ into the peptide embeddings to explore the latent space.
This noise $\delta$ is drawn from a uniform distribution $Z \sim \mathcal{U}(a,b)$, where $Z$ range in\texttt{[-1,1]}. To ensure the noise is not neutralized by normalization modules, we add a random shift $\beta$, sampled from a uniform distribution within the range \texttt{[-1,1]}, to the noise. This adjustment maintains the effectiveness of the noise and facilitates latent space exploration. The noise can be represent by: $\delta = \mathcal{U}(-1, 1)^{m \times n} + \beta$, where $\beta=\mathcal{U}(-1, 1)$, $m \times n$ represent the size of embedding matrix $f(x_i)$.

With lower noise levels, there is a higher likelihood of encountering sequences identical to the original. Conversely, excessively higher noise levels may produce sequences that diverge significantly from the original. To address this, we employ a strategy that gradually increases the noise level if no new sequences are discovered after a specified number of trials $\tau_{thresh}$. This adaptive approach strikes a balance between finding sequences that are similar to and those that are diverse from the original. The final noise applied to the embedding is given by: $\delta_\tau = (\alpha+\lfloor \tau/\tau_{thresh} \rfloor)* \delta)$ where $\alpha$ ($\alpha=0.5$ for initial noise) represents the noise level, and $\tau$ represent the current number of trials. For each trial, new noise and a random shift are generated. The trial thresholds $\tau_{thresh}$ vary for each model according to their computational time, ensuring trials are completed efficiently: $\tau_{thresh}=50$ for ProtT5 and $\tau_{thresh}=2000$ for ESM-2. This differentiation accounts for the different sizes of the models' parameters, with ProtT5 having approximately 2.8 billion parameters and ESM-2 having around 150 million parameters, thus requiring different amounts of time to complete one trial. Additionally, we found that ProtT5 usually identifies new peptide analogs in fewer trials compared to ESM-2 at the same level of noise. 

\subsection{Decoder}
The final step in our method involves transforming the noised embedding back into peptide sequences through function $g(\cdot)$. We employed different functions $g(\cdot)$ for ProtT5 and ESM-2 based on distinct architectures.

\subsubsection{ProtT5 Decoder}
We utilize a pre-trained decoder to project the transformed embeddings. The decoding process begins with the same initial token, \texttt{<\textbackslash{}pad>}, which initiates the transformation of the embedding back into a sequence. This allows for a straightforward and efficient decoding process, leveraging the capabilities of the pre-trained ProtT5 model.

\subsubsection{ESM-2 Decoder}
The ESM-2 does not have a decoder module, therefore, we trained our own decoder module. The ESM-2 decoder is designed to be symmetric with it encoder's architecture but with fewer layers and followed by a linear function $\hat{y} = h(z_{l-1}W)$ that maps the hidden state to the sequence token, where $z_{l-1}$ represent the hidden state from last layer of decoder module and $W$ is learnable weight matrix. Thefunction $h(\cdot)$ ensures that the decoder can effectively transform the embedding back into sequences. The ESM-2 decoder module is fine-tuned on the UniProtKB\cite{boutet2016uniprotkb} dataset using cross entropy loss. The parameters of the ESM-2 encoder remain frozen to ensure the embeddings are consistent with the original model. The frozen strategy maintains the integrity of the embeddings while allowing the decoder to learn the mapping from embeddings to sequences effectively.

\section{Data \& Experiment Setup}

\subsection{Data Source and Filtering}

In this study, we employed an open-source protein-ligand dataset: BioLip \cite{zhang2024biolip2} to test our method. BioLip contains a large volume of data, encompassing various ligand types. The recent update includes the sequences of peptides. The entire BioLip database contains 781,684 protein-ligand interactions, out of which 35,167 are protein-peptide interactions. After removing duplicate sequences, we are left with 9,027 unique peptide sequences. Further filtering out sequences containing non-standard amino acids reduces this number to 7,347 sequences. We then filter out sequences longer than 20 amino acids and shorter than 5 amino acids, resulting in a final dataset of 4,758 unique peptide sequences.

To train the decoder module for ESM-2, we used the UniProtKB/Swiss-Prot dataset which comprises 571,282 sequences\cite{boutet2016uniprotkb}. UniProtKB/Swiss-Prot (Universal Protein Resource Knowledgebase) is a comprehensive database that offers detailed information on protein sequences and functions. It is curated manually, focusing on proteins that are generally more well-researched. For sequences longer than 256, we truncated them from the head to ensure compatibility with the model. The number of sequences that are shorter than or equal to 256 is 245,539.

\subsection{Baseline Models}\label{sec:baseline}
In this study, we compare our proposed method with two baseline approaches. Random generated sequence and BLOSUM generated sequence. This comparison aims to highlight the efficacy and improvements offered by our proposed approach over traditional random sequence generation and selection based on global alignment metrics.

\subsubsection{Random Generated Sequence}
The first baseline method, random generated sequence, generates completely random sequences of the same length as the original sequences. This approach is commonly used in high-throughput lab experiments \ref{sec:lab_experiment_method} to test if some sequences display the desired properties. By using this method, we can evaluate whether our model's results are meaningful or if they could be attributed to random chance.

\subsubsection{BLOSUM Generated Sequence}
The second baseline involves a two-step process: initially generating 10,000 random sequences, followed by selecting the sequences with the highest global alignment scores using Needleman–Wunsch algorithm\cite{needleman1970general} with BLOSUM62 matrix\cite{henikoff1992amino}. 

\subsection{Evaluation Metrics}
To evaluate the similarity between original and generated peptide sequences, we use three different indicators: Morgan Fingerprints\cite{rogers2010extended}, RDKit Descriptors\cite{rdkit}, and QSAR descriptors\cite{althonos_peptides.py} based on peptide amino acid sequences. By employing these descriptors and their respective similarity measures, we ensure a comprehensive evaluation from structural, physico-chemical, and sequence-based perspectives. 

\subsubsection{Morgan Fingerprint}
We use the Morgan Fingerprint to represent the 2D structure of the molecules. To calculate the Morgan Fingerprint, we first transform the amino acid sequences into SMILES format using the molconvert tool from ChemAxon\cite{chemaxon_molconvert}. The Morgan Fingerprint captures the 2D structural features of the molecules by hashing these features into a fixed-size binary vector. For our evaluations, we choose a fingerprint size of 2048 bits. The similarity between two fingerprints is calculated using Tanimoto similarity, which measures the proportion of shared elements between two sets compared to their combined total.

\subsubsection{RDKit Descriptors}
RDKit\cite{rdkit} provides a variety of physico-chemical descriptors for molecules. To utilize RDKit for calculating these descriptors, we transform the amino acid sequences into SMILES format using the molconvert tool\cite{chemaxon_molconvert}. We use all the descriptors provided by RDKit to represent the peptides from a molecular aspect. The similarity between RDKit descriptors is calculated using cosine similarity. Before calculating the similarity, the descriptors are normalized to ensure accurate comparison.

\subsubsection{QSAR Descriptors}
QSAR (Quantitative Structure-Activity Relationship) descriptors for peptides encompass comprehensive properties and indices. These descriptors are calculated using the peptide.py package\cite{althonos_peptides.py} developed by European Molecular Biology Laboratory. Similar to RDKit descriptors, the QSAR descriptors are scaled and normalized before calculating similarity using cosine similarity.

\subsection{Comparative Analysis} \label{sec:MDexperimental}

To validate our proposed method, we applied it to peptide ligands of the TIGIT receptor, which have been identified through wet-lab experiments. We selected two peptide sequences: one with a strong affinity to the receptor as positive example and one with a weak affinity as negative example. For each peptide, we generated three similar sequences using our method. Please refer to the Supplementary Material for the exact sequences.

For a more accurate evaluation and further validation of the machine learning models, we employed MD Simulation. The process began with predicting the initial structure of each peptide sequence using AlphaFold2\cite{jumper2021highly}. The predicted initial structures were then placed with TIP3P water molecules into an MD system with a box size of 4 nm cubic box that mimics the experimental environment. The system was simulated for 1 ms with three repeats. We calculated the Root Mean Square Deviation (RMSD) between the last frames of the three repeats and then selected the repeat with the lowest RMSD compared to the other two. The structure of the TIGIT monomer was obtained from the Protein Data Bank (PDB ID: 3Q0H\cite{rcsb_3Q0H}). We created a system containing the TIGIT receptor and the peptide which was extracted from the peptide simulation for docking simulation, positioning the peptide around the pocket by a center of mass (COM) distance of 3 nm. The docking simulation was run for 500 ns with three repeats. The results were analyzed using three metrics: peptide RMSD, which shows the average positional deviation for each peptide residue during the simulation, indicating the stability of the peptide; the van der Waals (vdW) \& COM distance plot, which displays the vdW attractive energy and COM distance as a function of time, providing insights into the vdW energy at different distances between the pocket and the peptide during the simulation; and umbrella sampling\cite{torrie1977nonphysical}, which estimates the energy required to pull the peptide away from the bind TIGIT pocket, providing an estimate of the free energy (binding affinity) between the peptide and the TIGIT. Please refer to the supplementary material for more detailed implementation of the MD simulation.


\section{Result and Discussion}

\subsection{Overall Result}

We evaluated the performance of the proposed method against the two baseline models described in Section \ref{sec:baseline} in terms of average similarities for generating three, five, and 10 new sequences and the peptides with different length (shorter than 10, between 10 to 15, and longer than 15). As shown in Table ~\ref{tab:avg_similarities}, the proposed method outperforms all baseline models in terms of Morgan fingerprint, RDKit descriptor and sequence QSAR similarities. Specifically, ProtT5 exhibits higher average similarity for RDKit descriptor similarity indicating that ProtT5 may generate analogs with more similar physico-chemical properties, while ESM-2 shows higher average similarity for Morgan fingerprint and sequence QSAR similarities indicating that ESM-2 could generate analogs with more similar chemical structure and potentially similar bioactivities based on amino acid sequences. The BLOSUM and random generated sequence showed sub-optimal performance across all three similarity measures. Compare to random generate sequence, the BLOSUM generated sequence showed slightly better performance, however, it is still markedly underperformed compared to our proposed method. The underperformance of BLOSUM may be due to limitations in the BLOSUM matrix, which is based on a limited dataset and does not account for the context of amino acids, focusing only on position-specific substitutions\cite{eddy2004did}. Additionally, instances where miscalculated BLOSUM matrices outperformed correctly calculated ones suggest a gap between theoretical assumptions and practical performance\cite{styczynski2008blosum62}.

\setlength{\tabcolsep}{4pt} 
\begin{table}[htbp]
    \caption{Average Similarities}
    \begin{center}
\begin{tabular}{|l|l|r|r|r|r|}
\hline
\multicolumn{2}{|c|}{Method} & ProtT5          & ESM-2             & BLOSUM & Random \\ \hline
\multirow{6}{*}{\begin{tabular}[c]{@{}l@{}}Morgan \\ Fingerprint\end{tabular}} 
& 3 Sequences                      & 0.8155          & \textbf{0.8742} & 0.5572 & 0.3918 \\ \cline{2-6} 
& 5 Sequences                      & 0.7982          & \textbf{0.8745} & 0.5483 & 0.3915 \\ \cline{2-6} 
& 10 Sequences                     & 0.7697          & \textbf{0.8752} & 0.5359 & 0.3921 \\ \cline{2-6} 
& length\textless{}10              & 0.7228          & \textbf{0.8449} & 0.5393 & 0.3518 \\ \cline{2-6} 
& 10\textless{}length\textless{}15 & 0.7890          & \textbf{0.8889} & 0.5244 & 0.4000 \\ \cline{2-6} 
& 15\textless{}length              & 0.8350          & \textbf{0.9134} & 0.5408 & 0.4569 \\ \hline \hline

\multirow{6}{*}{\begin{tabular}[c]{@{}l@{}}RDKit \\ Descriptor\end{tabular}}   
& 3 Sequences                      & \textbf{0.9915} & 0.9497          & 0.9297 & 0.8874 \\ \cline{2-6} 
& 5 Sequences                      & \textbf{0.9894} & 0.9518          & 0.9281 & 0.8871 \\ \cline{2-6} 
& 10 Sequences                     & \textbf{0.9861} & 0.9550          & 0.9254 & 0.8872 \\ \cline{2-6} 
& length\textless{}10              & \textbf{0.9837} & 0.9456          & 0.9255 & 0.8761 \\ \cline{2-6} 
& 10\textless{}length\textless{}15 & \textbf{0.9869} & 0.9571          & 0.9193 & 0.8841 \\ \cline{2-6} 
& 15\textless{}length              & \textbf{0.9899} & 0.9704          & 0.9320 & 0.9105 \\ \hline \hline

\multirow{6}{*}{\begin{tabular}[c]{@{}l@{}}Sequence \\ QSAR\end{tabular}}      
& 3 Sequences                      & 0.9926          & \textbf{0.9961} & 0.9804 & 0.9603 \\ \cline{2-6} 
& 5 Sequences                      & 0.9911          & \textbf{0.9961} & 0.9797 & 0.9602 \\ \cline{2-6} 
& 10 Sequences                     & 0.9888          & \textbf{0.9960} & 0.9787 & 0.9602 \\ \cline{2-6} 
& length\textless{}10              & 0.9837          & \textbf{0.9939} & 0.9766 & 0.9516 \\ \cline{2-6} 
& 10\textless{}length\textless{}15 & 0.9915          & \textbf{0.9973} & 0.9792 & 0.9639 \\ \cline{2-6} 
& 15\textless{}length              & 0.9951          & \textbf{0.9986} & 0.9816 & 0.9713 \\ \hline
\end{tabular}
    \label{tab:avg_similarities}
    \end{center}
\textit{*Values are shown with four digits to highlight QSAR similarity differences.}

\textit{** For standard deviation values, please refer to Supplementary Material Table S1.}
\end{table}

\begin{figure}[htbp]
    \centering
    \includegraphics[width=\columnwidth]{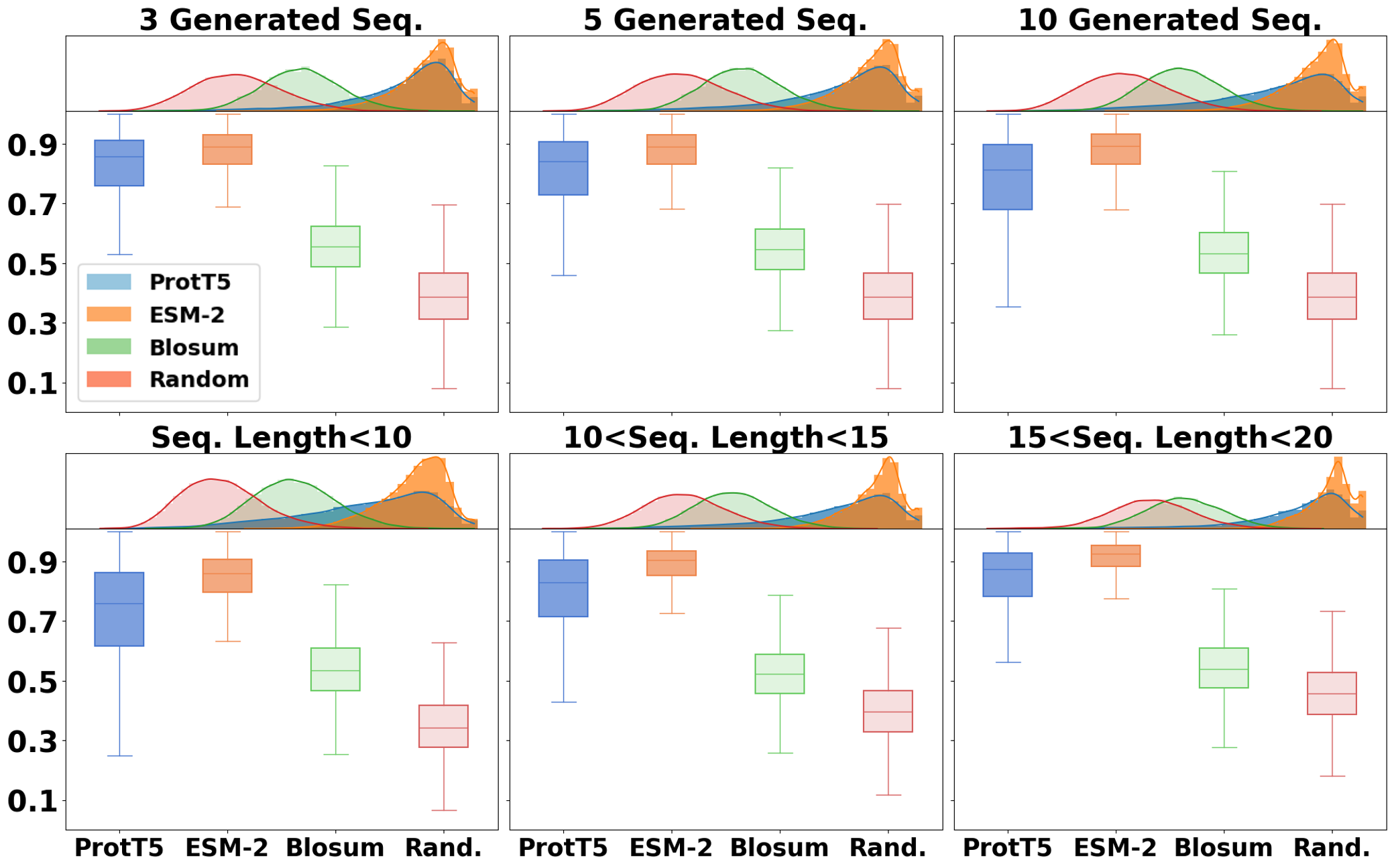}
    \caption{Morgan Fingerprint Similarities}
    \label{fig:morgan_fp_box}
\end{figure}
\begin{figure}[htbp]
    \centering
    \includegraphics[width=\columnwidth]{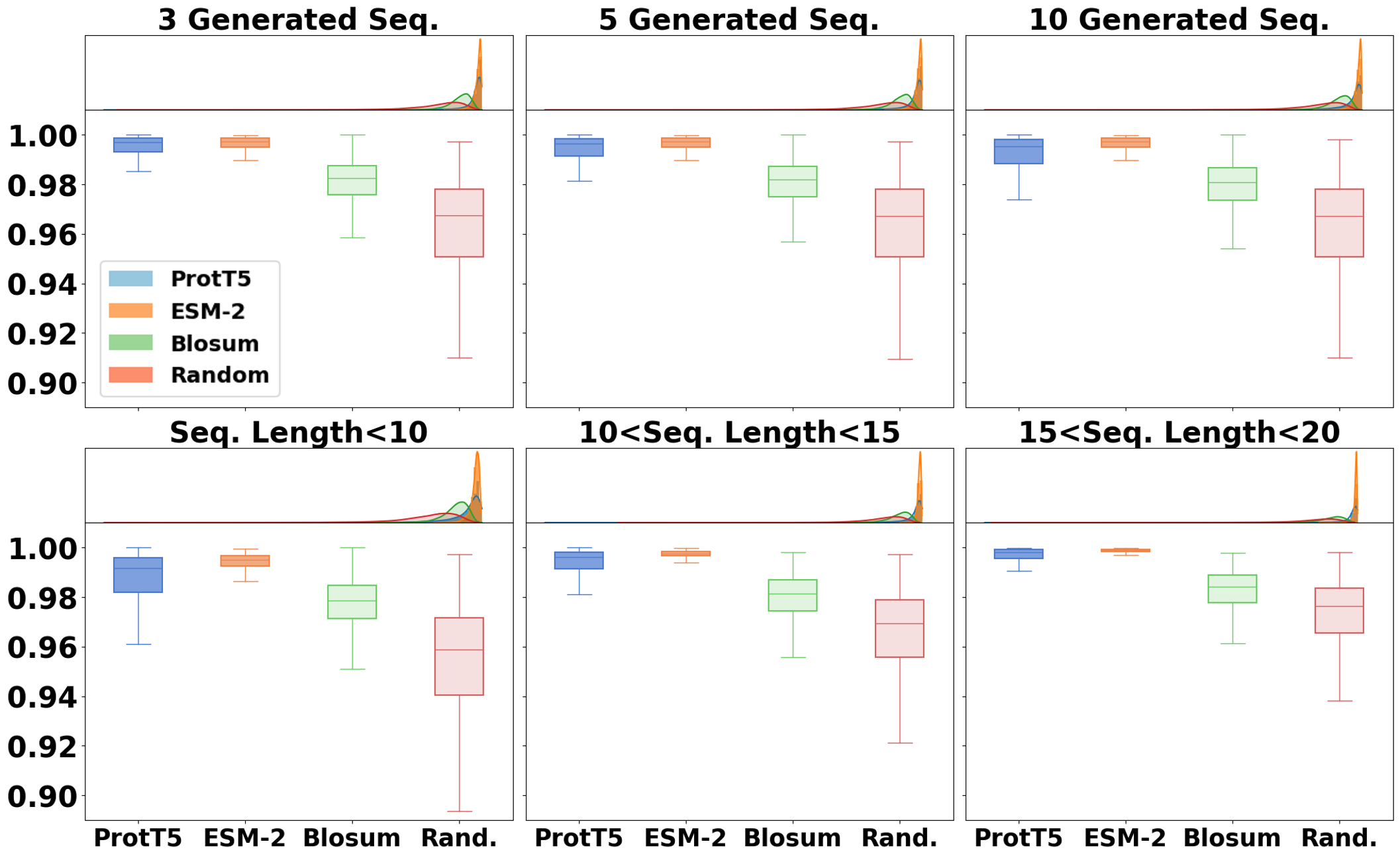}
    \caption{Sequences QSAR Similarities}
    \label{fig:qsar_box}
\end{figure}

As shown in Figure \ref{fig:morgan_fp_box}, \ref{fig:qsar_box}, for Morgan fingerprint and sequence QSAR similarities, sequences generated by the ESM-2 model have the highest similarities to the original sequences, followed by those generated by ProtT5. There is a large gap between these and the similarities achieved by the BLOSUM method, with random sequences showing the lowest similarities. Specifically, the ESM-generated sequences consistently demonstrate higher similarity metrics, indicating their effectiveness in maintaining key sequence properties. 

\begin{figure}[htbp]
    \centering
    \includegraphics[width=\columnwidth]{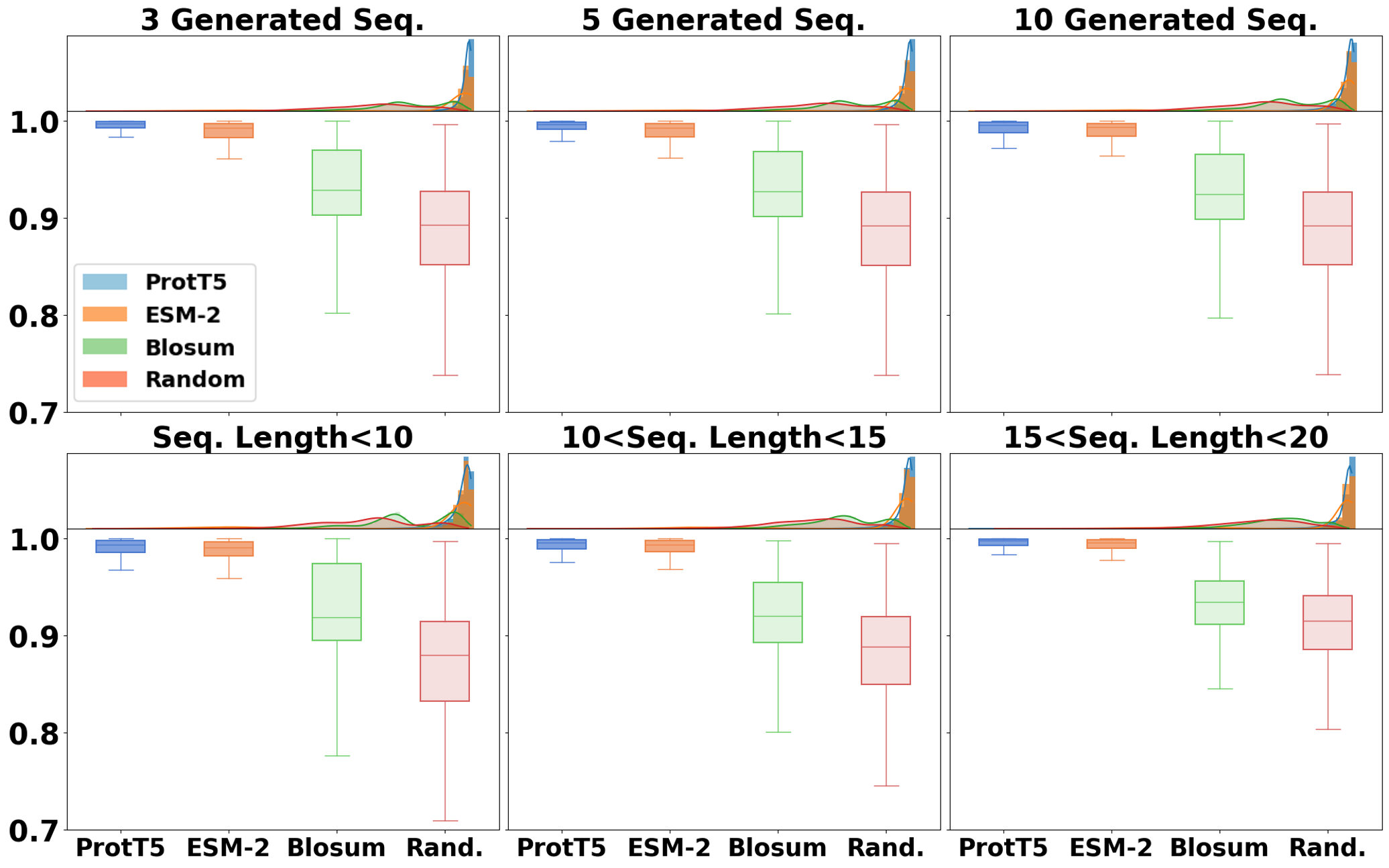}
    \caption{RDkit Descriptor Similarities}
    \label{fig:rdkit_box}
\end{figure}

In contrast, in Figure \ref{fig:rdkit_box}, for RDKit descriptor similarities, ProtT5-generated sequences outperform those generated by ESM-2. Both ProtT5 and ESM-2 models outperform the BLOSUM and random methods, suggesting that our approach yields peptides with more desirable molecular properties. 

We observed that the similarity distributions for all metrics are unimodal, regardless of sequence length or generation number. These results suggest that our method remains stable across different sequence lengths and maintains good performance even as the generation number increases. 
We also examined the impact of the number of generated sequences on performance, as shown in the Figure \ref{fig:morgan_fp_box}, \ref{fig:qsar_box}, \ref{fig:rdkit_box}, comparing sets of 3, 5, and 10 sequences. The results indicate a more pronounced decline in performance for ProtT5 as the number of sequences increases compared to ESM-2, suggesting that ProtT5 may be less stable when generating larger sets of sequences. Additionally, we analyzed the performance based on sequence length, categorizing sequences into three groups: shorter than 10, between 10 and 15, and longer than 15 amino acids. The findings reveal that longer sequences tend to have higher performance, even in baseline methods. This could be due to the larger search space available for longer sequences, which facilitates the generation of more similar peptides.

\begin{figure}[htbp]
    \centering
    \includegraphics[width=\columnwidth]{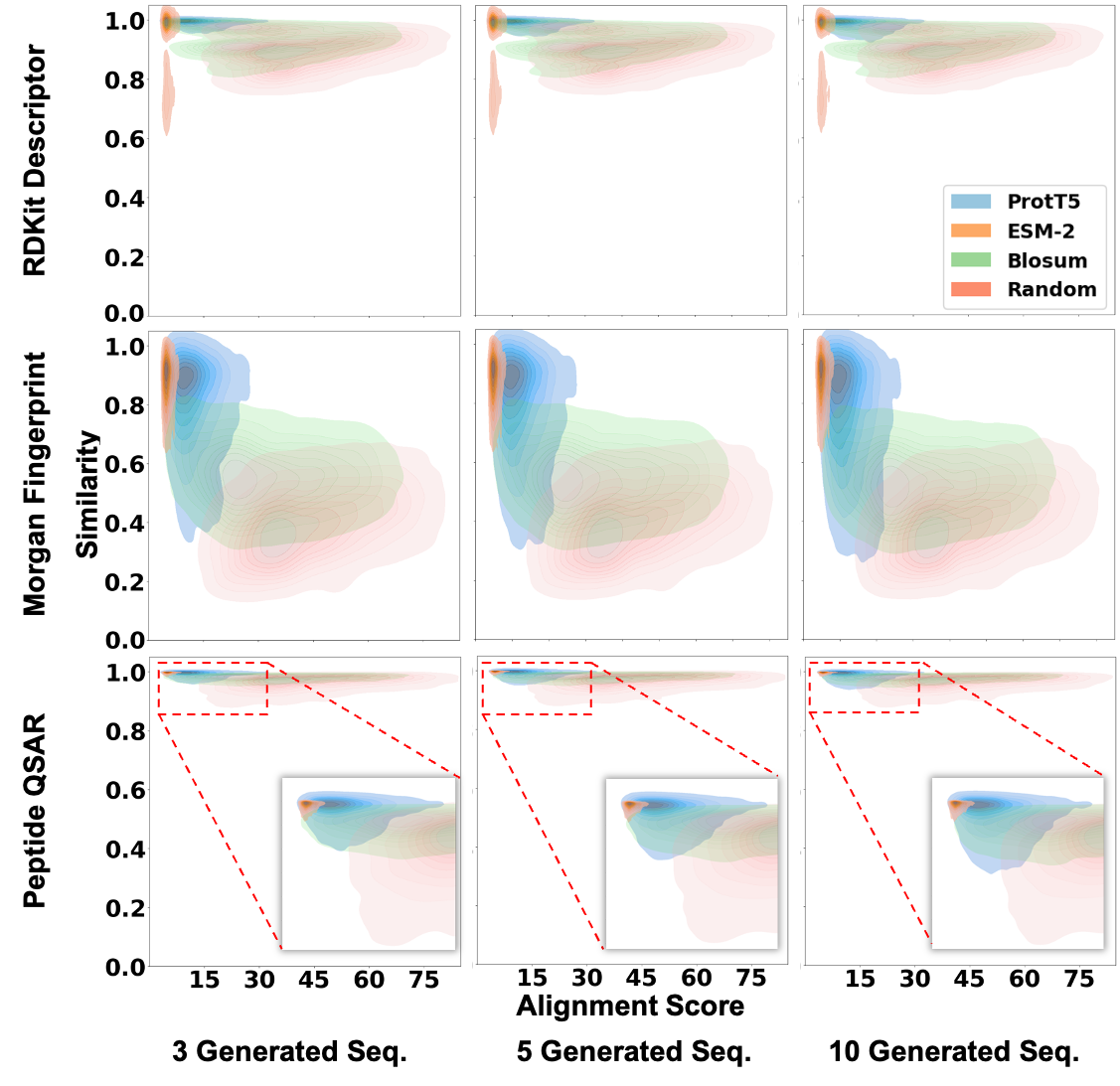}
    \caption{Similarities \& Alignment Score Difference}
    \label{fig:density_plot}
\end{figure}

To further understand the effectiveness of our method, we compared the distribution of similarities with alignment score differences using the BLOSUM matrix. The density plot Figure \ref{fig:density_plot} shows that our method can identify sequences with high property similarities while having significant differences in amino acid sequences compared to the baseline methods. This indicates that our approach can explore a diverse sequence space while maintaining crucial properties. When comparing ESM and ProtT5, ProtT5 shows more potential to search a larger sequence space with high similarity. On the other hand, ESM generally generates sequences with higher property similarities compared to ProtT5 but searches within a relatively smaller sequence space. Interestingly, in the RDKit descriptor similarities density plot, ESM occasionally generates sequences with lower similarity than the baseline methods. However, this phenomenon is rare and not significant enough to affect the overall performance.

\subsection{Physics Modeling Validation}

To further validate our proposed method, we evaluated the performance using physics model --- MD simulation --- with wet-lab experiments generated sequences. The generated sequence is obtained from ProtT5 due to it's ability to search for a broader range of sequences while maintaining high property similarity, making it a suitable choice for further exploration and validation in experimental settings.
\begin{figure}[htbp]
    \centering
    \includegraphics[width=0.8\columnwidth]{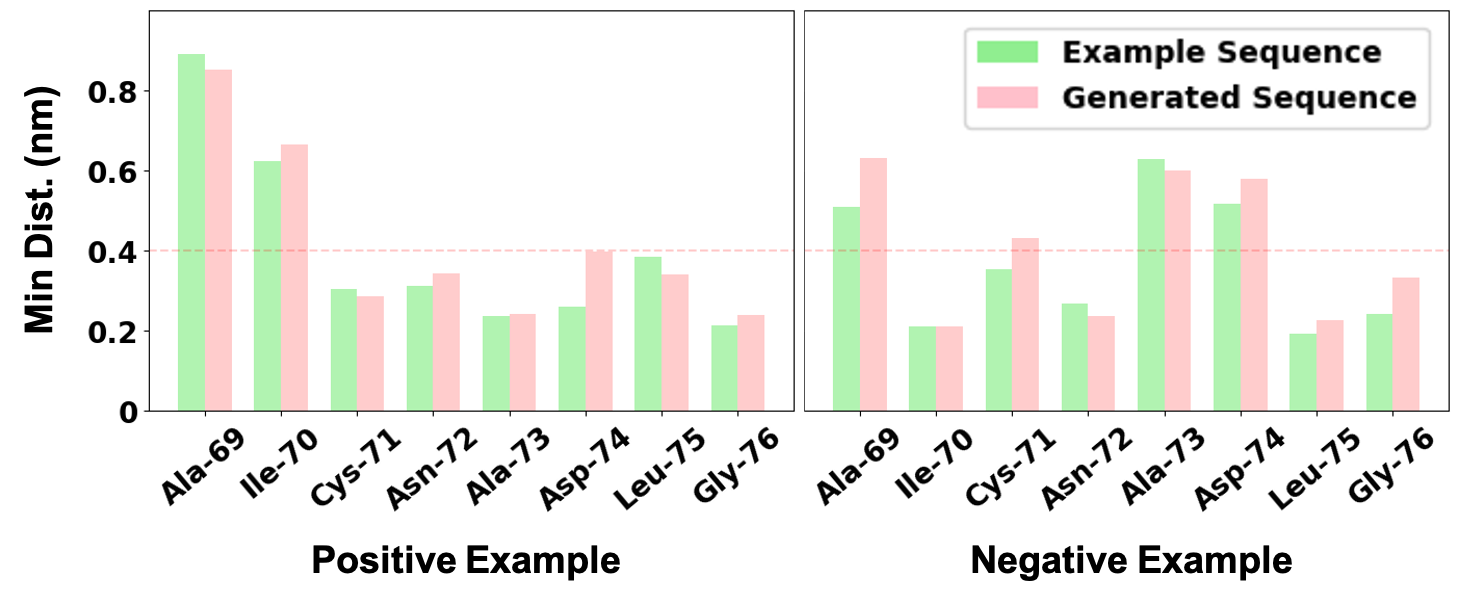}
    \caption{Minimum Distance to TIGIT Pocket Residues (Ala-67 to Gly-74)}
    \label{fig:minimum_pocket_dist}
\end{figure}
For our MD simulation analysis, we focus on the close interaction between TIGIT and peptide ligands, comparing positive and negative (the explanation of those pairs is detailed in section ~\ref{sec:MDexperimental}) examples. Among the three generated sequences, we selected the one with interactions most similar to the original peptide for further analysis. As shown in Figure \ref{fig:minimum_pocket_dist}, the minimum distance between pocket residues and the peptide exhibits similar behavior for both the positive and negative examples. We consider there to be an interaction between the peptide and a pocket residue if the minimum distance between them is less than 0.4 nm\cite{kumar2002close}. In the positive example, all pocket residues that interact with the example peptide also interact with the generated peptide. In the negative example, of the five residues interacting with the example peptide, only Cys-69 differs between the example and generated sequences. This difference results from the strict 0.4 nm cut-off, with the minimum distance difference being just 0.08 nm. Detailed residue-to-residue distances are included in Supplementary Material Figures S3 and S4.

\begin{figure}[htbp]
    \centering
    \includegraphics[width=\columnwidth]{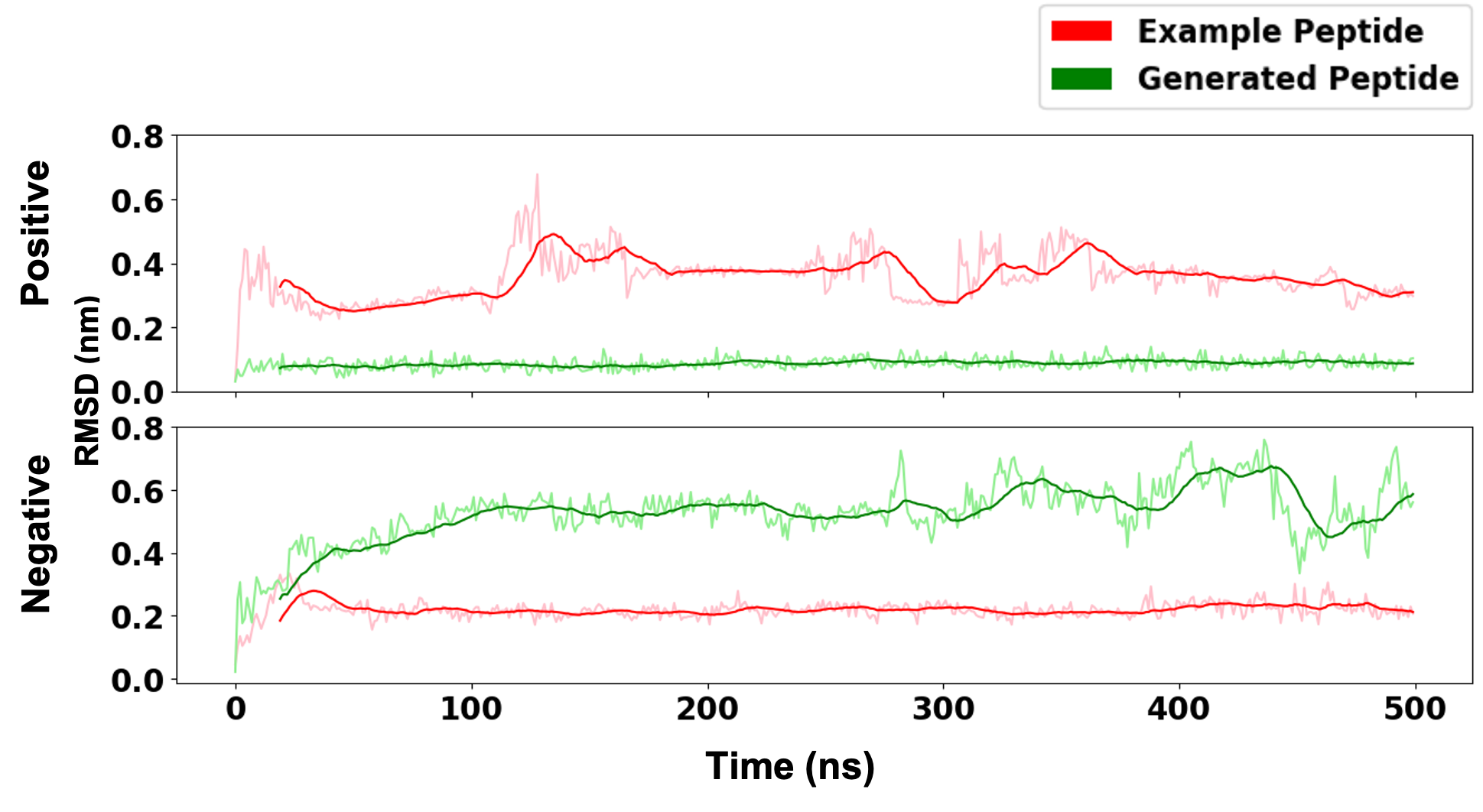}
    \caption{Peptide Root Mean Square Deviation}
    \label{fig:rmsd}
\end{figure}

\begin{figure}[htbp]
    \centering
    \includegraphics[width=\columnwidth]{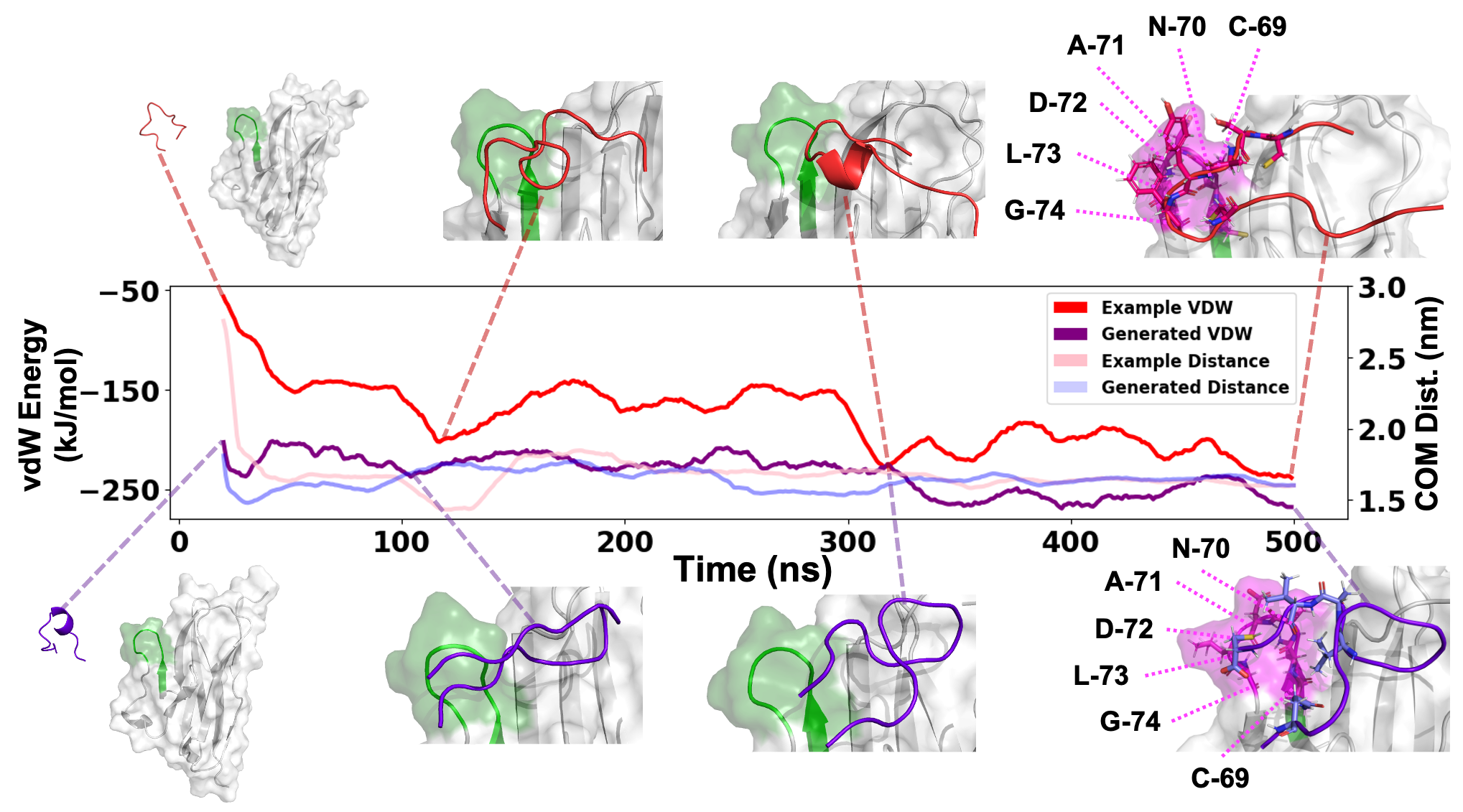}
    \caption{vdW \& COM distance - Positive Example. The first frame, showing vdW with a zero value, is cropped. Pocket residues have interaction with generated peptide are from Cys-69 to Gly-74, which are exactly same as the positive example peptide.}
    \label{fig:3p5_vdw_dist}
\end{figure}

\begin{figure}[htbp]
    \centering
    \includegraphics[width=\columnwidth]{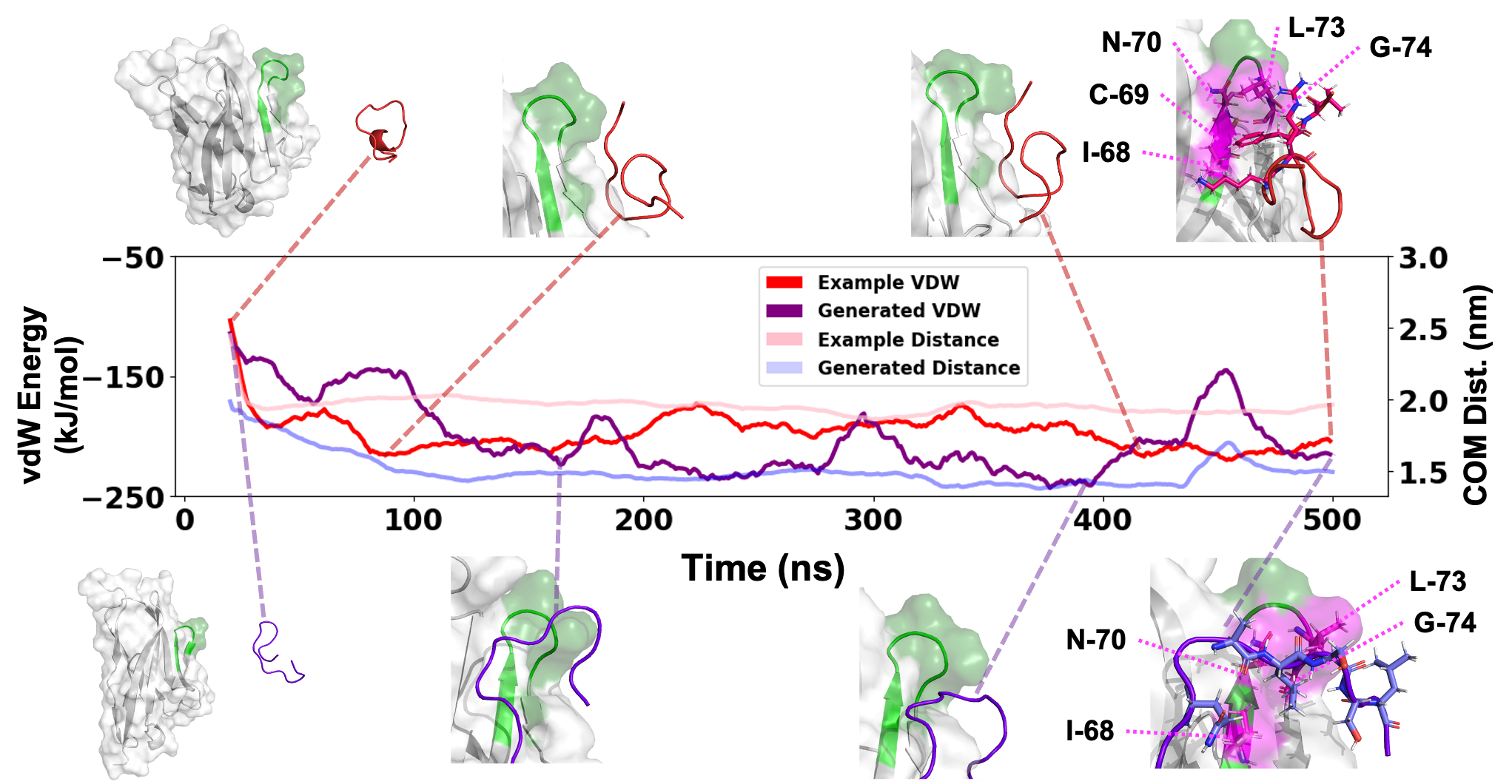}
    \caption{vdW \& COM distance - Negative Example.The first frame, showing vdW with a zero value, is cropped. Pocket residues have interaction with generated peptide are Ile-68, Asn-70, Leu-73, Gly-74, which are highly overlap with the negative example peptide.}
    \label{fig:22-1-1_vdw_dist}
\end{figure}

Based on the vdW \& COM distance plots (Figure \ref{fig:3p5_vdw_dist} and Figure \ref{fig:22-1-1_vdw_dist}), both positive and negative generated sequences show slightly higher vdW energy compared to the original sequences (15.8 kJ/mol for positive and 17.9 kJ/mol for negative). Additionally, for the negative example, the generated sequence exhibits a much lower COM distance (0.5 nm), indicating closer interaction with the pocket compared to the original negative peptide. This suggests that the generated sequence might bind more tightly to the pocket, potentially leading to improved function and higher efficacy.

In Figure \ref{fig:rmsd}, stability analysis through RMSD shows that the generated peptide in the positive example is more stable than the example peptide, with an average RMSD of 0.08 nm compared to 0.35 nm. Additionally, the generated peptide exhibits less fluctuation. Comparing this plot with Figure \ref{fig:3p5_vdw_dist}, we observe that the peptide quickly begins strongly interacting with the receptor and maintains its structure throughout the simulation. In the negative example, the generated peptide is less stable than the example peptide, with an average RMSD of 0.53 nm compared to 0.22 nm. Comparing this plot with Figure \ref{fig:22-1-1_vdw_dist}, around 400 ns, the peptide’s structure changes, causing the RMSD to increase as it slightly shifts from the interacting residues. It then engages with another set of pocket residues, leading to a subsequent drop in RMSD.

To evaluate the overall affinity, we utilized umbrella sampling. The results pf free energy, as shown in Figure S1 and Figure S2 in the Supplementary Material, reveal that in the positive example, the generated peptide demonstrates comparable affinity to the original peptide (52.00 kJ/mol for the generated peptide versus 58.11 kJ/mol for the original peptide). In the negative example, the generated peptide shows a much higher overall affinity (59.81 kJ/mol) compared to the original peptide (31.85 kJ/mol). These findings showcase the ability of our model not only to generate peptide analogs with similar behavior but also to potentially improve the affinity of existing sequences.

\section{Conclusion}
In this research, we have addressed the challenge of generating peptides with desired properties by proposing a novel method that efficiently produces peptide analogs. Traditional approaches in this domain often require large amounts of data, which can be a significant limitation. Our proposed method leverages the inherent capabilities of autoencoder models to explore the protein embedding space, relying solely on pre-trained protein language models. This allows our approach to generate new peptides using only a single sequence of interest, without the necessity for additional sequences with known properties or structures. Our results demonstrate that the proposed method significantly outperforms baseline models across three different similarity indicators.  To validate the robustness of our approach, we employed MD simulations on positive and negative examples of TIGIT inhibitors identified through wet lab experiments. These simulations revealed that our method successfully identified peptide analogs exhibiting behavior similar to the original positive and negative examples. 
Our findings suggest that the proposed method can significantly accelerate the peptide screening process by narrowing the search space. Future work will focus on testing our method in actual wet lab experiments to further validate its effectiveness.

\bibliographystyle{ieeetr}
\bibliography{ref}

\vspace{12pt}

\end{document}